\documentclass[12pt]{JHEP3}

\pdfoutput=1

\usepackage{amssymb,amsmath,amsfonts}
\usepackage{cite,slashed}
\usepackage{epsfig,subfigure}

\newcommand{\beqs}{\begin{equation*}}
\newcommand{\beq}{\begin{equation}}

\newcommand{\eeqs}{\end{equation*}}
\newcommand{\eeq}{\end{equation}}

\newcommand{\beqas}{\begin{eqnarray*}}
\newcommand{\beqa}{\begin{eqnarray}}

\newcommand{\eeqas}{\end{eqnarray*}}
\newcommand{\eeqa}{\end{eqnarray}}

\newcommand{\eq}[2]{\begin{equation} #1 \label{#2} \end{equation}}

\newcommand{\al}{\alpha}
\newcommand{\be}{\beta}
\newcommand{\ga}{\gamma}
\newcommand{\de}{\delta}

\newcommand{\ka}{\kappa}
\newcommand{\la}{\lambda}
\newcommand{\si}{\sigma}

\newcommand{\Om}{\Omega}

\newcommand{\La}{\Lambda}

\newcommand{\blist}{\begin{itemize}}

\newcommand{\elist}{\end{itemize}}

\providecommand{\href}[2]{#2}

\DeclareFontFamily{OT1}{rsfs}{}
\DeclareFontShape{OT1}{rsfs}{m}{n}{ <-7> rsfs5 <7-10> rsfs7 <10->rsfs10}{} 
\DeclareMathAlphabet{\mycal}{OT1}{rsfs}{m}{n}

\DeclareMathOperator{\extdm}{d}
\newcommand{\extd}{\extdm \!}

\newcommand{\Ric}{\slashed{R}}

\title{Generalised massive gravity one-loop partition function and AdS/(L)CFT}

\author{Mario Bertin$^a$, Daniel Grumiller$^b$, Dmitri Vassilevich$^{a,c}$ and Thomas Zojer$^b$\\
          $^a$~CMCC, Universidade Federal do ABC,\\
          Rua Santa Ad\'elia, 166,
          Santo Andr\'e, SP Brazil\\
          $^b$~Institute for Theoretical Physics, 
          Vienna University of Technology,\\
          Wiedner Hauptstr. 8--10/136,
          A-1040 Vienna, Austria\\
          $^c$~Department of Theoretical Physics, St Petersburg State University,\\
          St Petersburg, Russia\\
          Email: \email{mcbertin@gmail.com, grumil@hep.itp.tuwien.ac.at, dvassil@gmail.com, zojer@hep.itp.tuwien.ac.at}}

\abstract{
The graviton 1-loop partition function is calculated for Euclidean generalised massive gravity (GMG) using AdS heat kernel techniques.
We find that the results fit perfectly into the AdS/(L)CFT picture.
Conformal Chern--Simons gravity, a singular limit of GMG, leads to an additional contribution in the 1-loop determinant from the conformal ghost.
We show that this contribution has a nice interpretation on the conformal field theory side in terms of a semi-classical null vector at level two descending from a primary with conformal weights $(3/2,\,-1/2)$.
}


\preprint{TUW--11--06}

\begin{document}

\section{Introduction}

In the context of the AdS/CFT correspondence it is of interest to compare Euclidean partition functions calculated on the gravity side with partition functions for CFTs conjectured to be dual to the corresponding gravitational theory. 
A comparison of this kind works well for quantum gravity in AdS$_3$, since the dual CFT is 2-dimensional, and a great deal is known about such CFTs and their partition functions.
However, beyond the values of the central charges \cite{Brown:1986nw} little is known about the dual CFT from the gravity side, though there have been interesting conjectures and results recently \cite{Witten:2007kt,Li:2008dq,Grumiller:2008qz,Maloney:2009ck}.
The calculation of the quantum gravity partition function sheds further light on the properties of the dual CFT.
This provides the rationale for the present work.

To get started we split the metric $g$ into an AdS$_3$ background $\bar g$ and fluctuations $h$.
On the gravity side the full partition function $Z$ consists (at least) of two parts, a classical contribution $Z_{\rm cl}$ sensitive to the background $\bar g$ only, and a 1-loop contribution $Z_{\textrm{1-loop}}$ sensitive also to the fluctuations $h$.
\begin{equation}
Z = Z_{\textrm{cl}}\cdot Z_{\textrm{1-loop}}
\label{eq:intro1}
\end{equation}
For Einstein gravity the 1-loop partition function $Z_{\textrm{\tiny E}}$ was calculated efficiently using AdS heat kernel techniques \cite{Giombi:2008vd,David:2009xg}.
(For earlier papers and further references see \cite{Camporesi:1990wm,Camporesi:1994ga,Mann:1996ze,Bytsenko:1994bc}; for a review on heat kernel methods and $\zeta$-function renormalisation see \cite{Vassilevich:2003xt}.)
\eq{
Z_{\textrm{\tiny E}} = \prod_{n=2}^\infty \frac{1}{|1-q^n|^2} = Z_\Om  
}{eq:intro4}
Here $q=e^{i\tau}$, where $\tau=\tau_1+i\tau_2$, and $\tau_1$, $\tau_2$ correspond to angular potential $\theta$ and inverse temperature $\beta$, respectively.
More precisely, we assume that the background geometry $\bar g$ is thermal Euclidean AdS$_3$, which is the $M_{0,1}$ geometry in the notation of \cite{Maloney:2007ud}, with toric topology. 
Note that the full 1-loop partition function is a sum over partition functions $Z_{c,d}(\tau)$ that can be obtained from $Z_E=Z_{0,1}(\tau)$ by modular transformations \cite{Maloney:2007ud}.
In the present work we consider exclusively the $M_{0,1}$ geometry and its contribution to the partition function.
On the CFT side the result \eqref{eq:intro4} has a simple interpretation: it is just the partition function $Z_\Om$ of the Virasoro vacuum representation.
This is a basic example for a consistency check between the partition function calculated on the gravity side and the one expected from a CFT.
 
It was argued that for 3-dimensional Einstein gravity and chiral gravity \cite{Li:2008dq} the result \eqref{eq:intro1} coincides with the exact result \cite{Maloney:2007ud,Maloney:2009ck}.
This may or may not be true for more general theories of 3-dimensional gravity, like generalised massive gravity (GMG), whose action is given by \cite{Bergshoeff:2009hq}\footnote{An explanation of our notation is postponed to the beginning of section \ref{sec:2}.} 
 \begin{equation}\label{eq:intro2}
  S_{\textrm{\tiny GMG}} =  \frac{1}{\kappa^2}\int\extd^3x\left[ \sqrt{-g}\left(\sigma R
  +\frac{1}{m^2}\left(R^{\mu\nu}R_{\mu\nu}-\frac{3}{8}R^2\right)-2\lambda m^2\right)
  +\frac{1}{2\mu}{\cal L}_{\rm CS}\right]\;,
 \end{equation}
with the gravitational Chern--Simons term
 \begin{equation}
  {\cal L}_{\rm CS}  =  \varepsilon^{\mu\nu\rho}\left[
  \Gamma_{\mu\lambda}^{\sigma}\partial_{\nu}\Gamma_{\rho\sigma}^{\lambda}
  +\frac{2}{3}\Gamma_{\mu\lambda}^{\sigma}\Gamma_{\nu\kappa}^{\lambda}
  \Gamma_{\rho\sigma}^{\kappa}\right]\;.
\label{eq:intro3}
\end{equation}    
In any case, the 1-loop partition function certainly encodes crucial information about the gravity theory and its CFT dual, and like in the Einstein gravity example above it may be used for consistency checks of the AdS/CFT correspondence.

As another example let us consider the special case $m\to\infty$ of GMG, also known as topologically massive gravity (TMG) \cite{Deser:1982vy,Deser:1982wh,Deser:1982sv}.
The 1-loop partition function for TMG was calculated recently \cite{Gaberdiel:2010xv}, in the framework outlined above.
At the critical point $\mu\ell=1$ the 1-loop partition function reads
\begin{equation}
Z_{\textrm{TMG}}\Big|_{\mu\ell=1} = \prod\limits_{n=2}^\infty \frac{1}{|1-q^n|^2}\prod\limits_{m=2}^\infty\prod\limits_{\bar m=0}^\infty \frac{1}{1-q^m\bar q^{\bar m}} \,.
\label{eq:intro5}
\end{equation}
The result \eqref{eq:intro5} agrees with what one would expect from a log conformal field theory of the type proposed in \cite{Grumiller:2008qz}. 
It can be reformulated as follows.
\eq{
Z_{\textrm{TMG}}\Big|_{\mu\ell=1} = Z_{\textrm{\tiny LCFT}}^{(0)} + \sum_{h,\bar{h}} N_{h,\bar{h}}\, q^h \bar{q}^{\bar{h}} \, \prod_{n=1}^{\infty} \frac{1}{|1-q^n|^2} 
}{eq:intro7}
Here $Z_{\textrm{\tiny LCFT}}^{(0)}$ is the contribution to the log CFT partition function that takes into account the Virasoro descendants of the vacuum, as well as the Virasoro descendants of the log operator.
Thus, taking into account only single-particle log-excitations there is perfect agreement between the gravitational and the log CFT partition functions. 
However, $Z_{\textrm{\tiny LCFT}}^{(0)}$  does not take into account multi-particle log-excitations, which are contained in the second term on the right hand side of the result \eqref{eq:intro7} for the TMG 1-loop partition function.
This term describes the character of the $(h,\bar{h})$ representation of the Virasoro algebra, and $N_{h,\bar{h}}$ is the multiplicity with which this representation occurs. 
For consistency, all multiplicity coefficients $N_{h,\bar h}$ must be non-negative integers, and this indeed turned out to be the case.
The calculations performed in \cite{Gaberdiel:2010xv} therefore gave strong support to the idea that the dual of TMG at the critical point is a log CFT \cite{Grumiller:2008qz} (see also \cite{Maloney:2009ck}).  

Let us now come back to GMG.
For AdS boundary conditions and unity AdS radius the central charges of the dual CFT are given by
\eq{
  c_{R,\,L} = \frac{3}{2G}\,\left(\sigma+\frac{1}{2m^2}\pm \frac{1}{\mu}\right)\,. 
}{eq:intro9}
GMG is not only technically more complicated than TMG, but also exhibits novel features, like the possibility for multiple degeneration of modes or partial masslessness, with corresponding consequences for the dual CFT (see \cite{Grumiller:2010tj} and references therein). 
For instance, at the critical point (note that $\si^2=1$),
\eq{
m^2=2\mu=\frac32\,\si\,,
}{eq:intro8}
the left central charge vanishes, and the two massive graviton excitations degenerate with each other. 
Consequently, the theory was conjectured to be dual to a log CFT with higher rank Jordan cell \cite{Grumiller:2010tj}.
It is one of the aims of our work to test this conjecture, and more generally, to get insight into the nature of possible CFT duals at various values of the coupling constants.
 
Guided by the considerations above, in this paper we calculate the Euclidean 1-loop partition function for GMG,
\eq{
Z_{\textrm{\tiny GMG}} = \int{\cal D}h_{\mu\nu} \times [V_{\rm Diff}]^{-1}\times \exp{\big(-\de^{(2)}S_{\textrm{\tiny GMG}}\big)}\,,
}{eq:intro6}
where $\de^{(2)}$ denotes the second variation of the GMG action \eqref{eq:intro2},
and $V_{\rm Diff}$ is an (infinite) volume of the diffeomorphism group.
Instead of introducing a gauge-fixing term into the action, which may be inconvenient
in higher derivative theories, we take the same route as was taken in \cite{Gaberdiel:2010xv}
in the case of TMG. We explicitly separate gauge modes in $h_{\mu\nu}$, so that
the integration over these modes cancels the gauge group volume. This method is
especially effective on constant curvature spaces, see \cite{Vassilevich:1992rk,Mottola:1995sj}.
We then compare at special points in parameter space with the partition function of conjectured CFT duals.

This paper is organised as follows: 
In Section \ref{sec:2} we calculate the 1-loop partition function for GMG.
In Section \ref{sec:3} we consider various special cases and compare with CFT partition functions, including the intriguing case of Conformal Chern--Simons Gravity.
In Section \ref{sec:4} we summarise and address generalisations to extended massive gravity theories consistent with a holographic $c$-theorem.
In appendix \ref{app:phase} we derive a consistency relation for the phase of the 1-loop determinant.
In appendix \ref{app:A} we list the first ${\cal O}($hundred$)$ multiplicity coefficients for various special cases and display the typical combinatorial counting argument.
In appendix \ref{app:neg} we address Conformal Chern--Simons Gravity at negative temperature.

\section{One-loop partition function}\label{sec:2}

Before starting we mention our conventions.
The coupling constants are denoted as follows: 
$\si=\pm 1$ is the sign of the Einstein--Hilbert term,
$m^2$ is the coupling constant of new massive gravity,
$\mu$ is the Chern--Simons coupling constant, 
$\ka^2=16\pi G$ contains Newton's constant $G$.
We set $\ka=1$ unless stated otherwise.
We also set the AdS radius to unity, $\ell=1$, with no loss of generality. 
Our conventions for the GMG action are exactly as in \cite{Bergshoeff:2009aq}.
Our remaining conventions are exactly as in \cite{Gaberdiel:2010xv}.

\subsection{Generalised massive gravity}

Instead of working directly with the GMG action \eqref{eq:intro2} it is convenient to introduce an auxiliary field $f_{\mu\nu}$ \cite{Bergshoeff:2009aq}.  
 \begin{multline}
  S_{\textrm{\tiny GMG}}  =  \int\extd^3x\,\Big[ \sqrt{-g}\,\Big(\sigma R
  -2\lambda m^2+f^{\mu\nu}\big(R_{\mu\nu}-\frac12\,g_{\mu\nu}R\big)-\frac{1}{4}m^2\big(f^{\mu\nu}f_{\mu\nu}-f^2\big)\Big) \\
  +\frac{1}{2\mu}{\cal L}_{\rm CS}\Big]
\label{eq:gmg1}
 \end{multline}
We linearise now the fields.
 \begin{equation}
  g_{\mu\nu} \ = \ \bar{g}_{\mu\nu}+ h_{\mu\nu} \qquad
  f_{\mu\nu} \ = \ -\frac{1}{m^2}\,\left(\bar{g}_{\mu\nu}+ h_{\mu\nu}\right)
  -\frac{1}{m^2}k_{\mu\nu}
 \end{equation}
The fluctuation of $f_{\mu\nu}$ is defined such that $k_{\mu\nu}$ is gauge-invariant, whereas $h_{\mu\nu}$ transforms under gauge transformations in the usual way, 
  \begin{equation}
   \delta_{\xi} h_{\mu\nu} \ = \ \nabla_{\mu}\xi_{\nu}+\nabla_{\nu}\xi_{\mu}\,.
 \end{equation}
Let us continue the action \eqref{eq:gmg1} to Euclidean space.
It is enough to replace $\sqrt{-g}\to \sqrt{g}$ and $\varepsilon_{\mu}{}^{\rho\nu} \to i\varepsilon_{\mu}{}^{\rho\nu}$. 
The second variation of the Euclidean version of the GMG action \eqref{eq:gmg1} then reads
\begin{eqnarray}
&&\delta^{(2)}S_{\textrm{\tiny GMG}}  =  \int\extd^3x\sqrt{\bar{g}}\,\Big[ -\frac{1}{4}h^{\mu\nu} 
\left(\gamma 
+\frac{i}{2\mu} \widetilde{D} \right) L
   h_{\mu\nu} -
\frac{1}{2m^2}k^{\mu\nu} L h_{\mu\nu} \nonumber\\
&&\qquad\qquad\qquad  
   -\frac{1}{4m^2}\left(k^{\mu\nu}k_{\mu\nu}
-(k_{\mu\nu}\bar g^{\mu\nu})^2\right)\Big]\,.
  \phantom{\Bigl(} \label{E2S}
\end{eqnarray}
As in \cite{Gaberdiel:2010xv} we define 
\begin{eqnarray}
&&(\widetilde Dh)_{\mu\nu}= \varepsilon_\mu^{\ \, \rho\beta}\nabla_\rho h_{\nu\beta}
+ \varepsilon_\nu^{\ \, \rho\beta}\nabla_\rho h_{\mu\beta}\,,\label{tilD}\\
&&(Lh)_{\mu\nu}= -\nabla^2 h_{\mu\nu} -\nabla_\mu\nabla_\nu h + \nabla_\nu \nabla^\beta h_{\beta\mu}
+ \nabla_\mu \nabla^\beta h_{\beta\nu} - 2 h_{\mu\nu} \nonumber\\
&&\qquad\qquad - g_{\mu\nu} (\nabla_\rho\nabla_\sigma 
h^{\rho\sigma} - \nabla^2 h)  \label{defL}
\end{eqnarray}
 and additionally
\begin{equation}
\gamma=\sigma + \frac 1{2m^2}\,. \label{gamma}
\end{equation}

Let us York-decompose the tensor fields into transverse-traceless (TT), conformal and ``gauge'' parts.
\begin{eqnarray}
&&h_{\mu\nu}(h^{TT},h,\xi)=h_{\mu\nu}^{TT}+\frac 13 g_{\mu\nu} h + 2\nabla_{(\mu}\xi_{\nu)}\\
&&k_{\mu\nu}(k^{TT},k,v)= k_{\mu\nu}^{TT}+\frac 13 g_{\mu\nu} k + 2\nabla_{(\mu}v_{\nu)}\label{deco}
\end{eqnarray}
By definition $\nabla^\mu h^{TT}_{\mu\nu}=\nabla^\mu k^{TT}_{\mu\nu}=\bar g^{\mu\nu}h^{TT}_{\mu\nu}=\bar g^{\mu\nu}k^{TT}_{\mu\nu}=0$.
Due to gauge invariance the action does not depend on $\xi$.
The TT modes do not mix with the modes of other types. 
To remove the mixing between $h^{TT}$ and $k^{TT}$ we make the shift
\begin{equation}
\tilde k^{TT}_{\mu\nu}=k^{TT}_{\mu\nu}+Lh^{TT}_{\mu\nu}\,.\label{shTT}
\end{equation}
Then the action on TT fluctuations reads
\eq{
\delta^{(2)}S^{TT}_{\textrm{\tiny GMG}}=\int\extd^3x\sqrt{\bar{g}}\,\Big[
-\frac{m_1m_2}{4m^2} h^{TT\mu\nu} D^{m_1}D^{m_2}L\, h_{\mu\nu}^{TT}
-\frac 1{4m^2}\tilde k^{TT\mu\nu}\tilde k_{\mu\nu}^{TT} \Big]
}{eq:gmg2}
with
\begin{eqnarray}
&&(D^{m_{1,2}}h)_{\mu\nu} = h_{\mu\nu} + \frac i {2m_{1,2}} \,(\widetilde Dh)_{\mu\nu} \label{STT}\\
&&m_{1,2}=\frac{m^2}{2\mu} \pm \sqrt{ \frac 12 -\sigma m^2 +
\frac {m^4}{4\mu^2} }
\end{eqnarray}
where we used that on the TT modes $\widetilde D^2=4L-4$. 
Note also the useful relation $Lh_{\mu\nu}^{TT}=(-\nabla^2-2)h_{\mu\nu}^{TT}$. 

Let us now consider the vector modes. 
Since the action does not depend on $\xi$ we only need to consider $v$.
It is convenient to decompose $v$ into a transverse (T) and an exact contribution.
\begin{equation}
v_\mu=v_\mu^T +\partial_\mu u \qquad  \label{vTu}
\end{equation}
By definition $\nabla^\mu v_\mu^T=0$.
The T excitations do not mix with the other fields. 
The corresponding quadratic part of the action reads
\begin{equation}
\delta^{(2)}S^{T}_{\textrm{\tiny GMG}}=\int\extd^3x\sqrt{\bar{g}}\,\Big[ -\frac 1{2m^2} v^{T\mu} (-\nabla^2+2) v_\mu^T
\Big] \,.\label{S2T}
\end{equation}

The scalar sector consists of three fields, $h$, $k$ and $u$. 
After some algebra we obtain for $\ga\neq 0$
\begin{eqnarray}
&&\delta^{(2)}S^{S}_{\textrm{\tiny GMG}} 
=\int\extd^3x\sqrt{\bar{g}}\,\Big[ \frac{\gamma}{18} \tilde h (-\nabla^2 +3)
\tilde h + \frac 2{m^2} \tilde u\nabla^2 \tilde u \nonumber \\
&&\qquad\qquad\qquad\qquad  +
\frac 1{18m^2} \Big( 1-\frac 1{\gamma m^2} \Big)\, k(-\nabla^2+3)k\Big]\,,\label{S2S}
\end{eqnarray} 
where
\begin{equation}
\tilde h= h+ \frac 1{\gamma m^2} k \,,\qquad \tilde u=u+\frac 16 k\,.\label{shift}
\end{equation}
The case $\gamma=0$ will be considered below. 
For $\gamma m^2=1$ the field $k$ is a zero mode signalling an additional local symmetry in the action.
This feature corresponds to partial masslessness in the sense of Deser and Waldron \cite{Deser:1983mm,Deser:2001pe,Deser:2001us}, discussed in more detail in sections \ref{sec:2.2} and \ref{sec:3.5}. 
We assume for the time being $\ga m^2\neq 1$.

The relevant Jacobians were calculated in \cite{Gaberdiel:2010xv}:
\begin{eqnarray}
&&\mathcal{D}h_{\mu\nu}=Z_{\rm gh}\mathcal{D}h_{\mu\nu}^{TT}\,
\mathcal{D}\xi_\mu\, \mathcal{D}h\nonumber\\
&&\qquad Z_{\rm gh}=[\det (-\nabla^2+2)_1^T \det (-\nabla^2+3)_0]^{1/2}
\label{Zgh}\\
&&\mathcal{D}k_{\mu\nu}=J_2\, \mathcal{D}k_{\mu\nu}^{TT}\, \mathcal{D}v_\mu^T\,
\mathcal{D}u\, \mathcal{D}k\nonumber\\
&&\qquad J_2=[\det (-\nabla^2)_0\, \det (-\nabla^2+3)_0\, \det (-\nabla^2+2)_1^T]^{1/2}
\label{J2}
\end{eqnarray}
The subscripts near the operators denote the tensor rank of modes, while the superscripts signal restrictions on the modes. 
For example, the operator $(-\nabla^2+2)_1^T$ acts on transverse (T) vectors (1).
The shifts (\ref{shTT}) and (\ref{shift}) produce unite Jacobians. 
The 1-loop partition function reads
\begin{equation}
Z_{\rm GMG}=\int Z_{\rm gh}\mathcal{D}h_{\mu\nu}^{TT}\,
\mathcal{D}\xi_\mu\, \mathcal{D}\tilde h \cdot J_2\, \mathcal{D}\tilde k_{\mu\nu}^{TT}\, \mathcal{D}v_\mu^T\,
\mathcal{D}\tilde u\, \mathcal{D}k\, [V_{\rm Diff}]^{-1} e^{-\delta^{(2)}S_{\rm GMG}}\,. \label{intGMG}
\end{equation} 
Integration over $\xi$ cancels $[V_{\rm Diff}]^{-1}$, since $\delta^{(2)}S_{\rm GMG}$ 
does not depend on $\xi$.

The quadratic part of the action $\delta^{(2)}S_{\rm GMG}=\delta^{(2)}S^{TT}+
\delta^{(2)}S^{T}+\delta^{(2)}S^{S}$ is now diagonal in $h_{\mu\nu}^{TT}$,
$\tilde k_{\mu\nu}^{TT}$, $v_\mu^T$, $\tilde h$, $k$ and $\tilde u$. 
However, some of the kinetic terms have a wrong (negative) sign. This is similar
to negative sign of the kinetic for the trace part of the metric perturbations
in Euclidean Einstein gravity. Therefore, we use here the same remedy, namely the
Gibbons-Hawking-Perry rotation \cite{Gibbons:1978ac} 
of (some of) the fluctuations to imaginary values. More precisely, for positive
$m^2$ one should always rotate $\tilde k^{TT}$, $v^T$ and $\tilde u$. For negative
$\gamma$ one should also rotate $\tilde h$, while for $1/\gamma > m^2$ a rotation 
of $k$ is needed. All these rotations appear for auxiliary or otherwise non-propagating
fields and are presumably just gauge artifacts. As we shall see in a moment, all determinants
coming from rotated fields cancel out in the final result for the partition function,
which is a good consistency check. 
Indeed,  the integral over $h_{\mu\nu}^{TT}$, $\tilde k_{\mu\nu}^{TT}$, $v_\mu^T$, $\tilde h$, $k$ and $\tilde u$ gives
\eq{
\big[ \det (m_1D^{m_1}\cdot m_2D^{m_2}\cdot (-\nabla^2-2))_2^{TT} \cdot
\det (-\nabla^2+2)_1^T 
\cdot (\det (-\nabla^2+3)_0)^2 \cdot \det (-\nabla^2)_0 \big]^{-1/2}\,.
}{pathint}
Combining this result with the Jacobians (\ref{Zgh}) and (\ref{J2}) yields the 1-loop partition function for GMG,
\begin{equation}
Z_{\rm GMG}=\left[ \frac{\det( -\nabla^2+2)_1^T}{\det (m_1D^{m_1}
\cdot m_2D^{m_2}\cdot (-\nabla^2-2))_2^{TT}} \right]^{1/2} \,.\label{ZGMG}
\end{equation}

Let us consider now the case $\gamma=0$. 
The action for TT tensors and T vector modes is regular at $\gamma=0$. 
In the scalar sector we have
\begin{equation}
\delta^{(2)}S^{S}=\int\extd^3x\sqrt{\bar{g}}\,\Big[ \frac 1{9m^2} 
h(-\nabla^2+3)k +\frac 1{6m^2} k^2 
 +\frac {2}{3m^2} k \nabla^2 u +\frac 2{m^2} u\nabla^2 u
\Big] \,.\label{gamma0}
\end{equation}
First, we calculate the integral over $h$ which results in $\delta(k)\, \det (-\nabla^2 +3)_0^{-1}$. 
Then, the integration over $k$ is done trivially, and the integration over $u$ produces $[\det (-\nabla^2)_0]^{-1/2}$. 
The total contribution from the scalar modes to the partition function is then exactly the same as for $\gamma\neq 0$.

Thus, for $\gamma m^2\neq 1$ our final result for the 1-loop partition function for GMG is given by \eqref{ZGMG}, which we can rewrite conveniently as
\eq{
Z_{\rm GMG}=Z_{\rm E} \cdot Z_{m_1} \cdot Z_{m_2}
}{eq:gmg3}
with $Z_E$ being the Einstein 1-loop partition function,
\eq{
Z_E = \left[ \frac{\det( -\nabla^2+2)_1^T}{\det(-\nabla^2-2)_2^{TT}} \right]^{1/2} 
}{eq:gmg4}
and
\eq{
Z_{m_{1,2}} = \left[\det (m_{1,2}D^{m_{1,2}})\right]^{-1/2}\,.
}{eq:gmg5}

To evaluate the partition function we now use the results of \cite{Giombi:2008vd,David:2009xg}, who calculated the Einstein part \eqref{eq:intro4}, and of \cite{Gaberdiel:2010xv}, who calculated the contribution $Z_{m_{1,2}}$ up to a phase.
\eq{
\ln{|Z_{m_{1,2}}|} = \sum_{n=1}^\infty |q|^{n(|m_{1,2}|-1)}\, \frac{q^{2n} +\bar q^{2n}}{2n\,|1-q^n|^2} 
}{eq:gmg9}
We are going to discuss the choice of the phase in the next section for each example separately. 
All these choices are compatible with the consistency relation \eqref{eq:ZbZbq} derived in appendix \ref{app:phase}.

\subsection{Conformal Chern--Simons gravity}\label{sec:2.2}

If $\gamma m^2=1$ one of the modes becomes partially massless \cite{Bergshoeff:2009aq}, which means that there is an additional gauge symmetry
\eq{
\de_{\Om} k_{\mu\nu} = 2\Om \bar g_{\mu\nu} - 2\nabla_\mu\nabla_\nu\Om
}{eq:gmg6}
corresponding to linearised Weyl rescalings.
However, this symmetry in general is an artifact of the linearisation and does not persist in the full theory \cite{Blagojevic:2011qc}.
The only exception arises when the action consists solely of the gravitational Chern--Simons term.
\eq{
S_{\textrm{\tiny CSG}} = \frac{k}{4\pi}\,\int\extd^3x\,{\cal L}_{CS}
}{eq:gmg7}
We call that theory Conformal Chern--Simons gravity (CSG).
It is invariant in the bulk under finite local Weyl rescalings of the metric
\eq{
g_{\mu\nu}\to e^{2\Om}\,g_{\mu\nu}\,.
}{eq:Weyl}
Formally, CSG arises as a (singular) limit of GMG, $\mu\to 0$, $\ka^2\mu = 2\pi/k$, with some finite $k$.
However, as we shall see momentarily, the 1-loop partition function of CSG does not arise as the corresponding limit of the 1-loop partition function of GMG \eqref{ZGMG}.

The second variation of the CSG action yields
\begin{equation}
\delta^{(2)}S_{\textrm{\tiny CSG}}=  \frac{k}{4\pi}\,\int\extd^3x\sqrt{\bar{g}}\,\Big[ -\frac{i}{4}h^{\mu\nu} \widetilde{D} L
   h_{\mu\nu}\Big] \,.\label{delSCS}
\end{equation}
Due to the additional gauge invariance \eqref{eq:Weyl} the path integral measure has to be divided by the corresponding gauge group volume $V_{\rm conf}$, so that the 1-loop partition function reads
\begin{equation}
Z_{\textrm{\tiny CSG}} = \int{\cal D}h_{\mu\nu} \cdot [V_{\rm Diff}]^{-1}\cdot
[V_{\rm conf}]^{-1}\cdot
\exp{\big(-\de^{(2)}S_{\textrm{\tiny CSG}}\big)}\,.
\end{equation} 
After performing the York decomposition (\ref{deco}) for $h_{\mu\nu}$ we immediately
see that $\de^{(2)}S_{\textrm{\tiny CSG}}$ does not depend on $\xi_\mu$ and $h$.
The integration over these two fields cancels out the gauge group volumes.
The final result for the CSG partition function, 
\begin{equation}
Z_{\rm CSG}=\left[ \frac{\det( -\nabla^2+2)_1^T \det(-\nabla^2+3)_0}{\det (
\widetilde{D}\cdot (-\nabla^2-2))_2^{TT}} \right]^{1/2} =
Z_{\rm E}\cdot Z_{m=0} \cdot Z_{\rm conf}\label{ZCS}\,,
\end{equation}
contains a contribution from the conformal ghost
\eq{
Z_{\rm conf}=[\det(-\nabla^2+3)_0]^{1/2}\,.
}{eq:gmg8}
The appearance of the ghost determinant \eqref{eq:gmg8} is a major qualitative difference to the GMG result \eqref{ZGMG} and the TMG result \cite{Gaberdiel:2010xv}.

To evaluate the contribution of the conformal ghost to the partition function we use the results and notations of \cite{David:2009xg}.
\begin{eqnarray}
&&\ln Z_{\rm conf}= \frac 12 \ln \det (-\nabla^2 +3)_0 \nonumber\\
&&\qquad =-\frac 12 \int\limits_0^\infty \frac{\extd t}t K^0(t)e^{-3t} \nonumber\\
&&\qquad =-\frac 12 \sum_{n=1}^\infty \frac{\tau_2}{4\sqrt{\pi}
\left\vert \sin \frac{n\tau}2 \right\vert^2} \int\limits_0^\infty
\frac{\extd t}{t^{3/2}} e^{-\frac{n^2\tau_2^2}{4t} -4t} \nonumber\\
&&\qquad =-\frac 12 \sum_{n=1}^\infty \frac{1}{2n
\left\vert \sin \frac{n\tau}2 \right\vert^2} e^{-2n\tau_2} \nonumber\\
&&\qquad =- \sum_{n=1}^\infty \frac{|q|^{3n}}{n \,|1-q^n|^2} \label{lZc}
\end{eqnarray}
We assumed above positive (inverse) temperature, $\tau_2>0$.
The case of negative temperature, $\tau_2<0$, is treated in appendix \ref{app:neg}.

There is still one issue left open, namely the phase in \eqref{eq:gmg9}.
\eq{
\ln{|Z_{m=0}|} = \sum_{n=1}^\infty |q|^{-n}\, \frac{q^{2n} +\bar q^{2n}}{2n\,|1-q^n|^2} 
}{eq:gmg11}
There are two natural choices for the phase.
Either
\eq{
\ln{Z_{m=0}} = \sum_{n=1}^\infty \frac{|q|^{-n}\, q^{2n}}{n\,|1-q^n|^2} 
}{eq:gmg12}
or
\eq{
\ln{Z_{m=0}^{\rm alt}} = \sum_{n=1}^\infty  \frac{|q|^{-n}\,\bar q^{2n}}{n\,|1-q^n|^2} \,.
}{eq:gmg13}
Both choices --- as well as the fact that there are two choices --- are compatible with the CFT interpretation presented in section \ref{sec:3.5}.
The rationale behind these two choices will be explained below Eq.~\eqref{eq:gmg31}. 
It is a peculiar feature of CSG that there are two natural choices rather than one. 
This feature originates from the fact that one can approach the limit $m_2\to 0$ in \eqref{eq:gmg5} from above [leading to \eqref{eq:gmg12}] or from below [leading to \eqref{eq:gmg13}]. 
With no loss of generality we pick the former.

Collecting all contributions to the 1-loop partition function of CSG \eqref{ZCS} and exploiting the series expansion of the log function,
\eq{
\sum_{n=1}^\infty\,\frac{1}{n}q^{3n/2+mn}\bar q^{-n/2+\bar m n} (1-\bar q^{2n}) = \ln{\frac{1-q^{m+3/2}\bar q^{\bar m+3/2}}{1-q^{m+3/2}\bar q^{\bar m-1/2}}}
}{eq:gmg14}
we finally establish
\eq{
Z_{\rm CSG}=\prod_{n=2}^\infty \frac{1}{|1-q^n|^2} \,\prod_{m = 0}^\infty \frac{1}{\big(1-q^{m+3/2}\bar q^{-1/2}\big)\big(1-q^{m+3/2}\bar q^{1/2}\big)}\,.
}{eq:gmg10}
Note that in the result \eqref{eq:gmg10} there is only an infinite product over $m$, but not over $\bar m$, since in the latter all but two terms cancel.
By virtue of the Jacobi triple product identity the CSG 1-loop partition function \eqref{eq:gmg10} can be presented alternatively as
\eq{
Z_{\rm CSG}=\frac{\big(1-q\big)\big(1-\bar q\big)\big(1-\sqrt{\frac{q}{\bar q}}\big)\big(1-\sqrt{q\bar q}\big)}{\vartheta_4\big(-\bar\tau/4,\,\sqrt{q}\big) \,\eta(-\bar\tau/2\pi)\,\bar q^{-1/24}} \,
}{eq:angelinajolie}
with Jacobi's theta function ($\bar q = \exp{(-i\bar\tau)}$)
\eq{
\vartheta_4\big(-\bar\tau/4,\,\sqrt{q}\big) = \sum\limits_{m=-\infty}^\infty(-1)^m\,q^{m^2/2}\,\bar q^{m/2}
}{eq:lalapetz}
and Dedekind's eta function
\eq{
\eta(-\bar\tau/2\pi) = \bar q^{1/24} \prod_{n=1}^\infty \big(1-\bar q^n\big)\,.
}{eq:eta}
\begin{figure}[t]
 \centering
 \subfigure[]{
  \includegraphics[height=6.5cm]{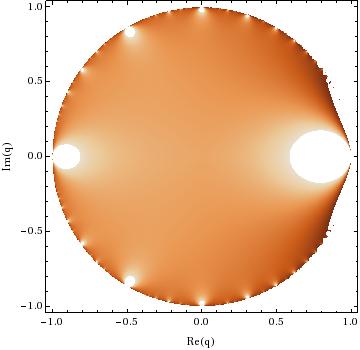}
   \label{fig:subfig1}
   }
 \subfigure[]{
  \includegraphics[height=6.5cm]{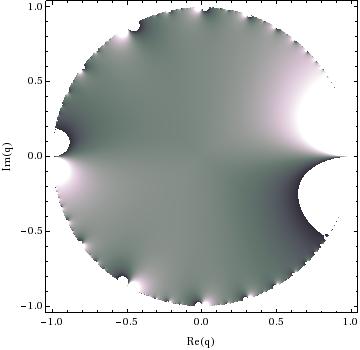}
   \label{fig:subfig2}
   }
 \caption{
  Real (a) and imaginary (b) parts of $\ln Z_{\rm CSG}$. The shading goes from darker (lower values) to brighter (higher values). The plots are cut off at large absolute values, so the white regions along the unit circle represent poles of either sign.}  \label{fig:Z}
\end{figure}

Real and imaginary parts of the logarithm of the CSG 1-loop partition function, $\ln Z_{\rm CSG}$, are displayed in figures~\ref{fig:subfig1} and \ref{fig:subfig2}.
The even poles at the roots of unity are clearly visible in the left plot. 
In this regard the real part of the partition function resembles the Einstein gravity partition function \eqref{eq:intro4}. 
The right plot, which would be trivial in Einstein gravity as the partition function \eqref{eq:intro4} is manifestly real, evidently is non-trivial and exhibits odd poles at roots of unity.

\section{Special cases and their CFT duals}\label{sec:3}

In this section we evaluate the GMG partition function \eqref{eq:gmg3}-\eqref{eq:gmg5} for special values of the coupling constants, and provide an educated guess for the phase not determined by the result \eqref{eq:gmg9}.
We then compare with the partition functions of conjectured CFT duals in order to support or falsify these conjectures.

In subsection \ref{sec:3.1} we discuss the critical point \eqref{eq:intro8}.
In subsection \ref{sec:3.2} we consider a critical line where $c_L=0$.
In subsection \ref{sec:3.3} we address a critical line where the two massive modes degenerate with each other and also mention the partially massless case $\ga m^2=1$.
In subsection \ref{sec:3.5} we study in great detail CSG \eqref{eq:gmg7}.

Before starting we mention two special cases that have been treated already in \cite{Gaberdiel:2010xv}: 
TMG ($m\to\infty$) and new massive gravity ($\mu\to\infty$). 
Both cases are recovered by taking the corresponding limits in the GMG partition function \eqref{eq:gmg3}-\eqref{eq:gmg5}.
This is of course expected and merely a consistency check on the correctness of the results above.

\subsection{Double log}\label{sec:3.1}

At the critical point \eqref{eq:intro8} both mass parameters degenerate with each other 
\begin{equation}
m_1=m_2=1 \qquad\mbox{and}\qquad \gamma m^2=2\,.\label{dchi}
\end{equation}
In addition, the massive modes both degenerate with the left-moving boundary graviton.
It was suggested in \cite{Grumiller:2010tj} that the dual CFT is a log CFT with the following structure.
The left central charge vanishes, $c_L=0$, whereas the right central charge is non-vanishing, $c_R=64\pi\si$.
The left-moving stress tensor $T(z)$ has two log partners, $t(z)$ and $\mathfrak{t}(z)$, satisfying
\eq{
 L_0\mathfrak{t} = 2\mathfrak{t} + t\qquad L_0t = 2t + T\qquad L_0 T = 2T \qquad L_1 \mathfrak{t} = L_1 t = L_1 T = 0 \,.
}{eq:gmg22}
Furthermore, the 2-point functions between $T$ and $\mathfrak{t}$ (and $t$ with itself) are given by $\langle T(z)\mathfrak{t}(w)\rangle = \langle t(z)t(w)\rangle = \frac{a_L}{2(z-w)^4}$.
This implies
\eq{
L_2\mathfrak{t} = a_L \Om\qquad a_L = c_R= 64\pi\si
}{eq:gmg23}
where $\Om$ is the ground state of the log CFT.
Here we have used $T=L_{-2}\Om$, which implies $L_2T=0$.
The 2-point function between $T$ and $t$ vanishes, and we have $L_2t=0$.
Moreover, $T$, $t$ and $\mathfrak{t}$ are annihilated by all positive $\bar L_n$ modes, as well as by $L_n$ with $n\geq 3$.
The angular momentum operator $L_0-\bar L_0$ is diagonalisable, which implies
\eq{
\bar L_0\mathfrak{t} = t\qquad \bar L_0 t = T \,.
}{eq:gmg24}
Finally, we have of course the property $\bar L_0 T = 0$. 
The structure of the low-lying states can therefore be summarised by the following diagram:
\begin{equation}
 \begin{picture}(180,160)(-10,-20)
      \put(-54,114){\vbox to 0pt
        {\vss\hbox to 0pt{\hss$\bullet$\hss}\vss}}
      \put(178,114){\vbox to 0pt
        {\vss\hbox to 0pt{\hss$\bullet$\hss}\vss}}
      \put(62,0){\vbox to 0pt
        {\vss\hbox to 0pt{\hss$\bullet$\hss}\vss}} 
      \put(62,114){\vbox to 0pt
        {\vss\hbox to 0pt{\hss$\bullet$\hss}\vss}}
      \put(173,114){\vector(-1,0){106}}
      \put(56,114){\vector(-1,0){104}}
      \put(57,3){\vector(-1,1){108}}
      \put(175,111){\vector(-1,-1){109}}
      \put(-59,124){$T$}
      \put(59,124){$t$}
      \put(177,124){$\mathfrak{t}$}
      \put(93,124){{\footnotesize $(L_0-2),\bar{L}_0$}}
      \put(-25,124){{\footnotesize $(L_0-2),\bar{L}_0$}}
      \put(54,-15){$\Omega$}
       \put(-17,47){{\footnotesize $L_{-2}$}}
        \put(127,47){{\footnotesize $L_{2}$}}
    \end{picture}
\end{equation}
In addition there is the right-moving stress energy tensor $\bar T = \bar L_{-2}\Om$ that satisfies the usual properties of the holomorphic flux component of a CFT stress energy tensor.
    
In order to determine the contribution of the above states to the partition function we can follow almost line-by-line the discussion in \cite{Gaberdiel:2010xv}.
We therefore immediately state the result for the partition function that counts the Virasoro descendants of the above states.
\eq{
Z_{\rm LCFT}^0 = Z_\Om + Z_t + Z_{\mathfrak{t}} = \prod_{n=2}^\infty \frac{1}{|1-q^n|^2}\,\Big(1+\frac{2q^2}{|1-q|^2}\Big)
}{eq:gmg25}
The result \eqref{eq:gmg25} differs from the TMG result in \cite{Gaberdiel:2010xv} by a factor of 2 in the second term in the brackets.
This is easily explained by the fact that here we have two log partners of $T$, whereas in the log CFT studied in \cite{Gaberdiel:2010xv} $T$ had only one log partner.

We turn now to the gravity side.
For $m_1=m_2=1$ the same choice for the phase as in \cite{Gaberdiel:2010xv} is natural, since again we have a degeneration with the left-moving boundary graviton.
With this choice we obtain from \eqref{eq:gmg3}-\eqref{eq:gmg5}
\eq{
Z_{\rm GMG}\Big|_{m_1=m_2=1} = \prod_{n=2}^\infty \frac{1}{|1-q^n|^2}\,\Big(\prod\limits_{m=2}^\infty\prod\limits_{\bar m=0}^\infty \frac{1}{1-q^m\bar q^{\bar m}} \Big)^2\,.
}{eq:gmg26}
We compare first the single-particle excitations on the gravity side
\eq{
\Big(\prod\limits_{m=2}^\infty\prod\limits_{\bar m=0}^\infty \frac{1}{1-q^m\bar q^{\bar m}} \Big)^2 = 1 + 2\sum_{m=2}^\infty\sum_{\bar m=0}^\infty q^m\bar q^{\bar m} + \textrm{multi-particle}
}{eq:gmg27}
and CFT side
\eq{
1+\frac{2q^2}{|1-q|^2}=1 + 2\sum_{m=2}^\infty\sum_{\bar m=0}^\infty q^m\bar q^{\bar m} \,.
}{eq:gmg28}
They match precisely.
We write now the gravity partition function \eqref{eq:gmg26} in the following suggestive form:
\eq{
Z_{\rm GMG}\Big|_{m_1=m_2=1} = Z_{\rm LCFT}^0 + \sum_{h,\,\bar h} N_{h,\,\bar h}\,q^h\bar q^{\bar h}\,\prod_{n=1}^\infty\frac{1}{|1-q^n|^2}
}{eq:gmg29}
As in \cite{Gaberdiel:2010xv} the last term describes the character of the $(h,\,\bar h)$ representation, and $N_{h,\,\bar h}$ is the multiplicity with which this representation occurs.
Again it is straightforward to prove by simple combinatorics that all multiplicity coefficients are non-negative integers, $N_{h,\,\bar h}\geq 0$.
We show this explicitly for the present case in appendix \ref{app:A}.

The GMG 1-loop calculation therefore gives strong support for the conjecture that GMG at the critical point \eqref{eq:intro4} is dual to a log CFT with the properties summarised above.

\subsection{Single log}\label{sec:3.2}

If one of the mass parameters equals to unity, say $m_1=1$, then one of the central charges vanishes, $c_L=0$.
The corresponding massive mode degenerates with the left-moving boundary graviton, and we should have a log CFT similar to the original proposal in the TMG context \cite{Grumiller:2008qz}.
Therefore, the contribution to the partition function on gravity and CFT sides from the log states, left- and right-moving boundary gravitons is precisely the same as in \cite{Gaberdiel:2010xv}.
This supports the conjecture that the dual CFT is a log CFT where the left-moving stress energy tensor acquires a log partner. 
However, as opposed to TMG there is an additional massive operator as long as $m_2^3\neq m_2$.
On the gravity side this leads to a contribution to the partition function of the form \eqref{eq:gmg9}.
It is not clear how to fix the phase in this expression, but we can provide a reasonable guess.
Namely, for positive $m_2$ we fix the phase such that
\eq{
\ln{Z_{m_2}} = \sum_{n=1}^\infty |q|^{n(m_2-1)}\, \frac{q^{2n}}{n\,|1-q^n|^2} \qquad m_2>0
}{eq:gmg30}
while for negative $m_2$ we fix the phase such that
\eq{
\ln{Z_{m_2}} = \sum_{n=1}^\infty |q|^{-n(m_2+1)}\, \frac{\bar q^{2n}}{n\,|1-q^n|^2} \qquad m_2<0 \,.
}{eq:gmg31}
The rationale behind these choices is that in the limits $m_2\to\pm 1$ the massive mode degenerates with the left- or right-moving boundary graviton, and the phase of the corresponding determinant is known. 
We assume that the phase does not change as long as $m_2$ does not change its sign.
In the limit $m_2\to 0$ (partial masslessness) there is an ambiguity that we have encountered already below \eqref{eq:gmg11}, and one can take either choice.
We address this case in more detail in subsections \ref{sec:3.4} and \ref{sec:3.5} below.

For specificity let us consider only the case of positive $m_2$.
If $m_2$ is an odd integer,%
\footnote{Amusingly, in TMG some odd integer values of the corresponding quantity, the Chern--Simons coupling $\mu$, play a special role already classically:
for $\mu=1$ one obtains log solutions \cite{Grumiller:2008qz}, for $\mu=3$ null warped solutions \cite{Anninos:2008fx} and for $\mu=5$ a special degenerate type of stationary, axi-symmetric solutions \cite{Ertl:2010dh}.
At 1-loop level the discussion is analog to here.} 
$m_2=2j+1$ with $j\geq 0$, then the partition function takes a particularly simple form,
\eq{
Z_{\rm GMG}\Big|_{m_1=0}^{m_2=2j+1} =\prod_{n=2}^\infty \frac{1}{|1-q^n|^2} \,\prod_{m = 2}^\infty\prod_{\bar m = 0}^\infty \frac{1}{1-q^{m}\bar q^{\bar m}}\,\prod_{l = j+2}^\infty\prod_{\bar l = j}^\infty \frac{1}{1-q^{l}\bar q^{\bar l}}\,.
}{eq:gmg32}
Analog to the discussion in sections \ref{sec:3.1} we can provide the CFT partition function that takes into account all Virasoro descendants of the vacuum $\Om$, the log quasi-primary $t$ and the massive primary $M$.
\eq{
Z^0_{\rm LCFT}(j) = Z_\Om + Z_t + Z_M(j) = \prod_{n=2}^\infty \frac{1}{|1-q^n|^2}\,\Big(1+\frac{q^2}{|1-q|^2}+\frac{q^{2+j}\bar q^{j}}{|1-q|^2}\Big)
}{eq:gmg37}
The partition functions match again precisely for single-particle excitations, and the full partition function can again be written as follows:
\eq{
Z_{\rm GMG}\Big|_{m_1=0}^{m_2=2j+1} = Z^0_{\rm LCFT}(j) + \sum_{h,\,\bar h} N_{h,\,\bar h}\,q^h\bar q^{\bar h}\,\prod_{n=1}^\infty\frac{1}{|1-q^n|^2}
}{eq:gmg38}
The positivity of the multiplicity coefficients $N_{h,\bar h}$ can again be shown using simple combinatorics.

Finally, note that requiring $m_2$ to be an odd integer quantises the right central charge as follows:
\eq{
c_R = \frac{12\si}{G}\,\frac{j+1}{4j+3} = 192\pi\si\,\frac{j+1}{4j+3}
}{eq:gmg33}
For these values of the central charge it is not unreasonable to expect a well-behaved log CFT dual, given that the matching of partition functions above worked so well.
It would be of interest to study the properties of these log CFTs in more detail.
We leave this to future work.
If $m_2$ takes arbitrary real values then the massive contribution to the partition function \eqref{eq:gmg30} in general will have a non-rational pre-factor in $q$ and $\bar q$.

\subsection{Massive log}\label{sec:3.3}\label{sec:3.4}

If $m_1=m_2$ then the massive modes degenerate with each other.
As long as $m_1^3\neq m_1$ we are in a generic situation.
Otherwise we recover either the double log case discussed in section \ref{sec:3.1} (for $m_1=\pm 1$) or the partially massless case mentioned below (for $m_1=0$).
Not much is known about the dual CFT, beyond the fact that it is a log CFT with non-vanishing central charges.
As before we focus here on values of the coupling constant where the partition function takes a particularly simple form ($j$ is some non-negative integer).
\eq{
m_1 = m_2 = \frac{m^2}{2\mu} = 2j+1 \qquad\Rightarrow\qquad m^2 = \si\big((2j+1)^2+\frac12\big)
}{eq:gmg35}
Choosing the phase as before we then obtain the partition function
\eq{
Z_{\rm GMG}\Big|_{m_1=m_2=2j+1} = \prod_{n=2}^\infty \frac{1}{|1-q^n|^2} \,\Big(\prod_{m = j+2}^\infty\prod_{\bar m = j}^\infty \frac{1}{1-q^{m}\bar q^{\bar m}}\Big)^2\,.
}{eq:gmg36}
Analog to the discussion in section \ref{sec:3.1} and \ref{sec:3.2} we can provide the CFT partition function that takes into account all Virasoro descendants of the vacuum $\Om$, the massive primary $M$ and its log partner $\mathfrak{m}$.
\eq{
Z^0_{\rm MLCFT}(j) = Z_\Om + Z_M(j) + Z_{\mathfrak{m}}(j)= \prod_{n=2}^\infty \frac{1}{|1-q^n|^2}\,\Big(1+\frac{2q^{2+j}\bar q^{j}}{|1-q|^2}\Big)
}{eq:gmg39}
The partition functions match again precisely for single-particle excitations, and the full partition function can again be written as follows:
\eq{
Z_{\rm GMG}\Big|_{m_1=m_2=2j+1} = Z^0_{\rm MLCFT}(j) + \sum_{h,\,\bar h} N_{h,\,\bar h}\,q^h\bar q^{\bar h}\,\prod_{n=1}^\infty\frac{1}{|1-q^n|^2}
}{eq:gmg40}
The positivity of the multiplicity coefficients $N_{h,\bar h}$ can again be shown using simple combinatorics.

We address now briefly the partially massless case $\ga m^2=1$.
Then at least one of the mass parameters vanishes (both vanish for partially massless gravity, $\mu\to\infty$, in which case the partially massless mode acquires a logarithmic partner \cite{Grumiller:2010tj}),
\eq{
m_{1,2} = \frac{\si}{4\mu} \pm \Big|\frac{\si}{4\mu}\Big|
}{eq:gmg34}
This means that there are necessarily contributions to the partition function that contain half integer powers in $q$ and $\bar q$, as evident from the expression \eqref{eq:gmg9} for $m_1\to 0$ or $m_2\to 0$.
However, as we have mentioned in section \ref{sec:2.2} the gauge symmetry that is present at the linearised level and that leads to a conformal ghost determinant ceases to exist in the full theory.
Therefore, it is not clear to us how seriously one should take the emergence of the conformal ghost determinant in GMG. 
We do no further dwell on this case.
Instead, we focus on CSG, where the gauge symmetry enhancement due to partial masslessness persists beyond linearisation, and the role of the conformal ghost determinant is well-understood.

\subsection{Conformal Chern--Simons gravity}\label{sec:3.5}

CSG \eqref{eq:gmg7} has rather interesting features on the gravity side \cite{CSGprep}.
As first pointed out in the context of partially massless gravity \cite{Grumiller:2010tj} there are three $SL(2,\mathbb{R})$ primaries with weights $(3/2,\,-1/2)$, $(-1/2,\,3/2)$ and $(3/2,\,3/2)$, corresponding to partially massless excitations.
One has to choose (by hand) if the $(3/2,\,-1/2)$ primary or the $(-1/2,\,3/2)$ primary is considered as normalisable \cite{CSGprep}.
This choice concurs with choosing either the phase \eqref{eq:gmg12} or \eqref{eq:gmg13}.
With no loss of generality we pick the $(3/2,\,-1/2)$ primary.
All its descendants then are also normalisable.
However, the $(-1/2,\,3/2)$ primary for consistency of the variational principle is then non-normalisable \cite{CSGprep}.
Finally, the $(3/2,\,3/2)$ primary (a scalar) has a fall-off behaviour that renders this mode and all its descendants pure gauge.
In the 1-loop calculation performed in section \ref{sec:2.2} the last feature is captured by the presence of the (scalar) conformal ghost determinant \eqref{eq:gmg8}.

On the CFT side there is a beautiful explanation for all these features.%
\footnote{We thank Matthias Gaberdiel and Niklas Johansson for discussions about this interpretation.
}
With the choice above in addition to the Virasoro descendants of the vacuum we have an operator with conformal weights $(3/2,\,-1/2)$.
We call it ``partially massless operator'' and denote the corresponding Virasoro primary state by $|P\rangle$.
As usual in the AdS/CFT context it is sourced by non-normalisable modes, which include the $(-1/2,\,3/2)$ primary.
The $(3/2,\,3/2)$ primary $|N\rangle=\bar L_{-1}^2|P\rangle$ is a level two descendant of the $(3/2,\,-1/2)$ primary.
Only $\bar L_2|N\rangle\neq 0$:
$L_n|N\rangle = 0$ for $n>0$ and $\bar L_n|N\rangle = 0$ for $n=1, n>2$.
We argue now that semi-classically $|N\rangle$ is actually a null vector.
To this end consider linear combinations of level two states $|\tilde N\rangle :=(\alpha \bar L_{-2}+\beta \bar L_{-1}^2)|P\rangle$, where we set $\be=1$ with no loss of generality.
Next, we require $|\tilde N\rangle$ to be a null vector.
\eq{
\langle\tilde N|\tilde N\rangle = \langle P| (\bar\alpha \bar L_2+\bar L_1^2) (\alpha \bar L_{-2}+\bar L_{-1}^2)|P\rangle = 0
}{eq:null}
The null condition \eqref{eq:null} holds if the state $|\tilde N\rangle$ is annihilated by all positive Virasoro generators. 
Requiring $\bar L_n |\tilde N\rangle=0$ for all positive $n$, in particular for $n=1,\,2$, yields the conditions $\al = -\frac43\, (\bar h+\frac12)$ and $\bar h=\frac{1}{16}\,\big(5-c\pm\sqrt{(1-c)(25-c)}\big)$, where $\bar h$ is the $\bar L_0$-weight that the primary $|P\rangle$ must have in order for $|\tilde N\rangle$ to be a null vector.
In the limit of large central charges $c$ these conditions establish (we discard another solution that would imply negative infinite weight)
\eq{
\alpha = \frac{6}{c} + {\cal O}(1/c^2) \qquad \bar h = -\frac12 - \frac{9}{2c} + {\cal O}(1/c^2)\,.
}{eq:nullprime}
Thus, in the semi-classical limit $c\to\infty$ a null vector at level two exists provided the weight $\bar h$ of the primary approaches $-1/2$.
This is precisely the weight of the primary $|P\rangle$ above.
Consistently, in the semi-classical limit we obtain 
\eq{
\lim_{c\to\infty}|\tilde N\rangle = |N\rangle
}{eq:nulllimit}
We call $|N\rangle$ a semi-classical null vector.

Thus, the Verma module generated by the partially massless operator is reducible as it contains a submodule generated by the semi-classical null vector $|N\rangle$.
The partition function $Z_{\rm CFT}^0$, defined as 
\eq{
Z_{\rm CFT}^0 = Z_\Om + Z_{pm}\,,
}{eq:gmg15}
takes into account the Virasoro descendants of the vacuum \eqref{eq:intro4} and the Virasoro descendants of the partially massless operator,
\eq{
Z_{pm} = q^{3/2}\bar q^{-1/2}\,\prod_{n=1}^\infty \frac{1}{|1-q^n|^2} - q^{3/2}\bar q^{3/2}\,\prod_{n=1}^\infty \frac{1}{|1-q^n|^2}\,.
}{eq:gmg17}
The second term in \eqref{eq:gmg17} is subtracted in order to eliminate the semi-classical null vector and its Virasoro descendants.
We thus obtain
\eq{
Z_{\rm CFT}^0 = \prod_{n=2}^\infty \frac{1}{|1-q^n|^2}\,\Big(1+\frac{q^{3/2}\bar q^{-1/2}(1-\bar q^2)}{|1-q|^2}\Big)\,.
}{eq:gmg18}

In the spirit of \eqref{eq:intro7} we consider the contribution of the partially massless states to the CSG partition function \eqref{eq:gmg10} and expand, keeping to leading order only the linear term from each denominator.
\begin{multline}
\prod_{m = 0}^\infty \frac{1}{\big(1-q^{m+3/2}\bar q^{-1/2}\big)\big(1-q^{m+3/2}\bar q^{1/2}\big)} = \\
1 + \sum_{m=0}^\infty q^{m+3/2}\bar q^{-1/2} + \sum_{m=0}^\infty q^{m+3/2}\bar q^{1/2}  + \textrm{multi-particle}
\label{eq:gmg20}
\end{multline}
Following \cite{Gaberdiel:2010xv} we interpret the terms listed explicitly on the right hand side of \eqref{eq:gmg20} as the single-particle partially massless excitations.
The gravity result \eqref{eq:gmg20} then matches perfectly the CFT result \eqref{eq:gmg18}, since
\eq{
1+\frac{q^{3/2}\bar q^{-1/2}(1-\bar q^2)}{|1-q|^2} = 1 + \sum_{m=0}^\infty q^{m+3/2}\bar q^{-1/2} + \sum_{m=0}^\infty q^{m+3/2}\bar q^{1/2}\,.
}{eq:gmg21} 
We can now present the result for the CSG partition function \eqref{eq:gmg10} in the following suggestive form.
\eq{
Z_{\rm CSG} = Z_{\rm CFT}^0 + \sum_{h,\,\bar h} N_{h,\,\bar h}\,q^h\bar q^{\bar h}\,\big(1-\bar q^2\big)\,\prod_{n=1}^\infty\frac{1}{|1-q^n|^2}
}{eq:gmg19}
The sum in \eqref{eq:gmg19} extends over all non-negative integer and half-integer values of $h$ and over all integer and half-integer values of $\bar h$.
It is crucial to include the factor $(1-\bar q^2)$ in the last term of \eqref{eq:gmg19} in order to get an irreducible character.%
\footnote{Indeed, if one would use instead the Virasoro character, like e.g.~in the result \eqref{eq:gmg29}, then the corresponding multiplicity coefficients would not be non-negative.
Instead, one would obtain the following relation, $N_{h,\,1/2+\bar h} + N_{h,\,1/2-\bar h} = 0$, and inequality, $N_{h,\,1/2+\bar h}\leq 0$, valid for $h$, $\bar h\geq 0$.
}
The multiplicity coefficients $N_{h,\bar h}$ can be non-zero only if the sum $h+\bar h$ is integer.
Thus, we have to calculate only $N_{n,\,m}$ and $N_{n+1/2,\,m+1/2}$ for integers $n$, $m$.
The first few coefficients are displayed in the last two tables in appendix \ref{app:A}.
The positivity of the remaining multiplicity coefficients $N_{h,\bar h}$ can again be shown using combinatorics. 

In the semi-classical limit $c\to\infty$ we have therefore excellent agreement between the partition functions on the gravity side and the CFT side.
If the null condition \eqref{eq:null} persists beyond the semi-classical approximation the weight $\bar h$ must acquire an anomalous contribution given in \eqref{eq:nullprime}.

\section{Summary and generalisations to extended massive gravity}\label{sec:4}

In the present work we calculated the 1-loop partition function for generalised massive gravity (GMG) on the gravity side in Section \ref{sec:2}. 
In Section \ref{sec:3} we compared the results for the partition function with corresponding results on the CFT side.
Our results strongly support the conjecture that GMG at certain critical loci in parameter space is dual to specific log CFTs. 

Conformal Chern--Simons gravity (CSG) is a non-trivial limit of GMG with an additional gauge symmetry, namely bulk Weyl rescaling. 
Interestingly, in this case the contribution from the additional 1-loop determinant precisely cancels the contributions of descendants of the semi-classical null vector at level two. 
The resulting partition function can be expressed in terms of $\vartheta$- and $\eta$-functions \eqref{eq:angelinajolie}.
The positivity of the multiplicity coefficients $N_{h,\bar h}$ in the expansion \eqref{eq:gmg19} provides fairly non-trivial support for a CFT interpretation with a reducible Verma module.

All of the CFTs mentioned above are non-unitary, because they contain either log operators or have negative central charge and/or states with negative conformal dimensions.
However, in CSG it might be possible to extract some unitary CFT, possibly by redefining $\bar L_n\to-\bar L_{-n}$, which essentially flips the sign of one of the central charges.
It could be interesting to pursue further the case of CSG, with particular focus on holography \cite{CSGprep}.

It is straightforward to generalise our results in three dimensions.
Paulos and Sinha put forward higher-derivative gravity theories that allow for a holographic $c$-theorem \cite{Sinha:2010ai,Paulos:2010ke,Sinha:2010pm}. 
These extended massive gravity theories have the action
\eq{
S = \frac{1}{\ka^2}\,\int\extd^3x\sqrt{-g}\,\big[\sigma R-2\La+\sum_{nmk}\,\la_{nmk} R^n R_{(2)}^m R_{(3)}^k +\frac{1}{2\mu}{\cal L}_{\rm CS}\Big]\,,
}{eq:conclusion1}
where the scalars $R_{(2)}=\Ric_{\mu\nu}\Ric^{\mu\nu}$ and $R_{(3)}=\Ric_{\mu\nu}\Ric^\mu_\al\Ric^{\nu\al}$ are quadratic and cubic curvature invariants constructed from the tracefree Ricci-tensor $\Ric_{\mu\nu}=R_{\mu\nu}-\frac13\,Rg_{\mu\nu}$. 
The coupling constants $\la_{nmk}$ are restricted by the existence of a holographic $c$-theorem. 
To quadratic order the action \eqref{eq:conclusion1} matches the action of GMG.
At the linearised level the equations of motion of these extended massive gravity theories take the same form as in GMG. 
They are of fourth order and, by fine-tuning of the coupling constants $\lambda_{nmk}$, allow the same kinds of degenerations and limiting cases as GMG. 
Similar remarks apply to Born--Infeld gravity \cite{Gullu:2010pc,Gullu:2010st}, which can also be expanded in the form \eqref{eq:conclusion1}.
The 1-loop partition functions of all these extended massive gravity theories should be equivalent to the partition function of GMG calculated in the present work.

\acknowledgments

We are grateful to Hamid Afshar, Branislav Cvetkovic, Sabine Ertl, Matthias Gaberdiel, Olaf Hohm and Niklas Johansson for discussions and collaborations on related topics.
We additionally thank Radoslav Rashkov for discussions.

MB was supported by FAPESP.
DG and TZ were supported by the START project Y435-N16 of the Austrian Science Foundation (FWF). 
DV was supported in part by CNPq and FAPESP. 
DG is grateful for the hospitality at USP, UF ABC and IFT in Sa{\~o} Paulo, where most of this work was performed, and acknowledges financial support from FAPESP.
DV acknowledges financial support from the Erwin-Schr\"odinger Institute (ESI) during the workshop ``Gravity in three dimensions''.  

\paragraph{Note added}
Matthias Gaberdiel informed us that \cite{GGHR} observe a similar phenomenon as in section \ref{sec:3.5} for semi-classical ($c\to\infty$) null vectors in the $W_\infty$ algebra.

\begin{appendix}

\section{Restriction on the phase of the massive determinants}\label{app:phase}

Let us take the metric on Euclidean AdS, $\mathbb{H}_3$, with unit AdS radius in global coordinates ($w:=t+i\phi$)
\begin{equation}
\extd s^2 = \extd\rho^2 + \frac14\, \big(\extd w^2 + \extd\bar w^2\,\big) +\frac12\, \cosh{(2\rho)}\,\extd w \extd\bar w
\end{equation}
The transformation $w\to \bar w$ is an isometry that reverses the orientation. 
Under this transformation the operator $\widetilde D$ in \eqref{tilD} is mapped to $\widetilde D^c$.
Eigenmodes of $\widetilde D$ with eigenvalue $\lambda$,
\begin{equation}
\widetilde D h^{TT}_\lambda (w,\,\bar w,\,\rho)=\lambda  h^{TT}_\lambda (w,\,\bar w,\,\rho)
\end{equation}
are mapped to eigenmodes of $\widetilde D^c$ with eigenvalue $-\lambda$.
\begin{equation}
\widetilde D^c h^{TT}_\lambda (\bar w,\, w,\,\rho)=-\lambda  h^{TT}_\lambda (\bar w, \,w,\,\rho)
\end{equation}
Consider the operator $D^m$ acting on TT tensors. 
From the definition \eqref{STT} it is obvious that the transformation $w\to\bar w$ implies complex conjugation of the eigenvalues of eigenmodes of $D^m$.

To construct the manifold $M_{0,1}$ one has to identify the points $w$ and $qw$ on $\mathbb{H}_3$.
The transformation $w\to \bar w$ maps $q$ to $\bar q$. 
Therefore, we establish a consistency relation.
\begin{equation}
Z_m(q)=\bar Z_m (\bar q) \label{eq:ZbZbq}
\end{equation}
Equation \eqref{eq:ZbZbq} imposes restrictions on phases of determinants and particularly forbids contributions to $\ln Z_m$ of the type $i f(|q|)$ with some real function $f$.
All phases chosen in the main text are compatible with the consistency relation \eqref{eq:ZbZbq}. 

\section{Multiplicity coefficients}\label{app:A}

{\footnotesize
\TABULAR{|r|ccccccccccc|}{
\hline
$\bar h=$ & 0 & 1 & 2 & 3 & 4 & 5 & 6 & 7 & 8 & 9 & 10 \\ \hline
$h \leq 3$: & 0 & 0 & 0 & 0 & 0 & 0  & 0 & 0 & 0 & 0 & 0  \\
$h=4$: & 3 & 1 & 3 & 1 & 3 & 1 & 3 & 1 & 3 & 1 & 3 \\
$h=5$: &  1 & 3 & 1 & 3 & 1 & 3 & 1 & 3 & 1 & 3 & 1 \\
$h=6$: & 7 & 3 & 9 & 7 & 11 & 9 & 15 & 11 & 17 & 15 & 19 \\
$h=7$: & 3 & 9 & 9 & 13 & 15 & 19 & 19 & 25 & 25 & 29 & 31 \\
$h=8$: & 14 & 12 & 26 & 28 & 43 & 43 & 63 & 63 & 84 & 86 & 108 \\
$h=9$: & 10 & 22 & 32 & 50 & 61 & 85 & 101 & 127 & 146 & 180 & 200 \\
$h=10$:& 26 & 34 & 71 & 89 & 142 & 168 & 235 & 273 & 358 & 406 & 509 \\
\hline
}{Double log multiplicity coefficients $N_{h,\bar h}$ in \eqref{eq:gmg29} for $h, \bar h < 11$}
}
{\footnotesize
\TABULAR{|r|ccccccccccc|}{
\hline
$\bar h=$ & 0 & 1 & 2 & 3 & 4 & 5 & 6 & 7 & 8 & 9 & 10 \\ \hline
$h \leq 3$: & 0 & 0 & 0 & 0 & 0 & 0  & 0 & 0 & 0 & 0 & 0  \\
$h=4$: & 1 & 0 & 1 & 0 & 1 & 0 & 1 & 0 & 1 & 0 & 1 \\
$h=5$: & 0 & 2 & 1 & 2 & 1 & 2 & 1 & 2 & 1 & 2 & 1 \\
$h=6$: & 2 & 1 & 4 & 2 & 4 & 2 & 5 & 2 & 5 & 3 & 5 \\
$h=7$: & 0 & 4 & 3 & 6 & 5 & 8 & 6 & 10 & 8 & 11 & 10 \\
$h=8$: & 3 & 3 & 9 & 8 & 14 & 12 & 19 & 17 & 24 & 22 & 29 \\
$h=9$: & 1 & 7 & 9 & 18 & 18 & 28 & 30 & 39 & 41 & 53 & 54 \\
$h=10$:& 4 & 7 & 19 & 22 & 39 & 43 & 61 & 68 & 90 & 96 & 123 \\
\hline
}{Single log multiplicity coefficients $N_{h,\bar h}$ for $j=1$ in \eqref{eq:gmg38} for $h, \bar h < 11$}
}
{\footnotesize
\TABULAR{|r|ccccccccccc|}{
\hline
$\bar h=$ & 0 & 1 & 2 & 3 & 4 & 5 & 6 & 7 & 8 & 9 & 10 \\ \hline
$h \leq 5$: & 0 & 0 & 0 & 0 & 0 & 0  & 0 & 0 & 0 & 0 & 0  \\
$h=6$: & 0 & 0 & 3 & 1 & 3 & 1 & 3 & 1 & 3 & 1 & 3 \\
$h=7$: & 0 & 0 & 1 & 3 & 1 & 3 & 1 & 3 & 1 & 3 & 1 \\
$h=8$: & 0 & 0 & 3 & 1 & 3 & 1 & 3 & 1 & 3 & 1 & 3 \\
$h=9$: & 0 & 0 & 1 & 7 & 3 & 9 & 7 & 11 & 9 & 15 & 11 \\
$h=10$:& 0 & 0 & 3 & 3 & 9 & 9 & 13 & 15 & 19 & 19 & 25 \\
\hline
}{Massive log multiplicity coefficients $N_{h,\bar h}$ for $j=1$ in \eqref{eq:gmg40} for $h, \bar h < 11$}
}
{\footnotesize
\TABULAR{|r|cccccccccccccc|}{
\hline
$\bar h=$ & -3 & -2 & -1 & 0 & 1 & 2 & 3 & 4 & 5 & 6 & 7 & 8 & 9 & 10 \\ \hline
$h < 3$: &  0 & 0 & 0 & 0 & 0 & 0 & 0 & 0 & 0 & 0 & 0 & 0 & 0 & 0 \\
$h=3$: & 0 & 0 & 1 & 1 & 1 & 1 & 2 & 3 & 4 & 5 & 7 & 9 & 12 & 15 \\
$h=4$: & 0 & 0 & 1 & 2 & 1 & 1 & 3 & 4 & 5 & 7 & 9 & 12 & 16 & 20 \\
$h=5$:&  0 & 0 & 3 & 4 & 3 & 3 & 7 & 10 & 13 & 17 & 23 & 30 & 40 & 50 \\
$h=6$:&  0 & 1 & 5 & 8 & 6 & 7 & 14 & 20 & 26 & 35 & 46 & 61 & 80 & 102 \\
$h=7$:&  0 & 1 & 10 & 14 & 11 & 12 & 25 & 36 & 47 & 62 & 83 & 109 & 144 & 182 \\
$h=8$:&  0 & 3 & 16 & 25 & 19 & 22 & 44 & 63 & 82 & 110 & 145 & 192 & 252 & 321 \\
$h=9$:&  1 & 6 & 30 & 43 & 37 & 43 & 81 & 116 & 153 & 203 & 270 & 355 & 467 & 594 \\
$h=10$:& 1 & 12 & 47 & 71 & 60 & 72 & 132 & 190 & 250 & 334 & 441 & 583 & 763 & 976  \\
\hline
}{Integer CSG multiplicity coefficients $N_{h,\bar h}$ in \eqref{eq:gmg19} for $-4 < \bar h < 11$ and $h < 11$}
}
{\footnotesize
\TABULAR{|r|cccccccccccccc|}{
\hline
$2\bar h=$ & -7 & -5 & -3 & -1 & 1 & 3 & 5 & 7 & 9 & 11 & 13 & 15 & 17 & 19  \\ \hline
$2h < 9$: &  0 & 0 & 0 & 0 & 0 & 0 & 0 & 0 & 0 & 0 & 0 & 0 & 0 & 0  \\
$2h=9$: &  0 & 0 & 1 & 1 & 1 & 2 & 2 & 3 & 5 & 6 & 8 & 11 & 14 & 18  \\ 
$2h=11$: &  0 & 0 & 1 & 2 & 2 & 2 & 3 & 5 & 7 & 9 & 12 & 16 & 21 & 27  \\
$2h=13$:&  0 & 0 & 3 & 5 & 5 & 6 & 8 & 13 & 19 & 24 & 32 & 43 & 56 & 72  \\
$2h=15$:&  0 & 1 & 6 & 10 & 11 & 13 & 18 & 28 & 40 & 52 & 69 & 92 & 120 & 155  \\
$2h=17$:&  0 & 1 & 11 & 19 & 20 & 23 & 32 & 51 & 73 & 94 & 125 & 167 & 218 & 281  \\
$2h=19$:&  0 & 3 & 19 & 34 & 37 & 41 & 59 & 93 & 131 & 171 & 227 & 302 & 395 & 510  \\
$2h=21$:&  1 & 6 & 35 & 61 & 67 & 77 & 109 & 170 & 240 & 313 & 416 & 553 & 723 & 933  \\
\hline
}{$\frac12$-integer CSG multiplicity coefficients $N_{h,\bar h}$ in \eqref{eq:gmg19} for $-4 < \bar h < 10$ and $h < 11$}
}
In this appendix we list the first $\cal O($hundred$)$ multiplicity coefficients for various cases discussed in the main text, since in all the combinatorial proofs the first few coefficients have to be determined explicitly.
All the coefficients in these tables are non-negative integers.
The last two tables both refer to CSG, with the former displaying results for integer values of $(h,\,\bar h)$ and the latter displaying results for $\frac12$-integer values.

We provide now explicitly the combinatorial counting argument that proves non-negativity of all multiplicity coefficients for the double log case discussed in section \ref{sec:3.1}, following appendix B of \cite{Gaberdiel:2010xv}.
(The proof is analogous for the other cases discussed in this paper.)
The Fourier coefficients $B(h,\,\bar h)$ in the double expansion \eqref{eq:gmg27}, 
\eq{
D=\Big(\prod_{m=2}^\infty\prod_{\bar m=0}^\infty\frac{1}{1-q^m\bar q^{\bar m}}\Big)^2 = 1 + \sum_{h,\bar h} B(h,\,\bar h)\,q^h\,\bar q^{\bar h}\,,
}{eq:app1}
are manifestly non-negative integers, $B(h,\,\bar h)\geq 0$.
By construction, the Fourier coefficients $\tilde B$ of
\eq{
\tilde D = D(1-q)(1-\bar q) = 1 + \sum_{h,\bar h} \tilde B(h,\,\bar h)\,q^h\,\bar q^{\bar h}
}{eq:app2}
satisfy the relation
\eq{
\tilde B(h,\,\bar h) = B(h,\,\bar h) - B(h-1,\,\bar h) - B(h,\,\bar h-1) + B(h-1,\,\bar h-1)\,.
}{eq:app3}
The inequality 
\eq{
B(h,\,\bar h)  + B(h-1,\,\bar h-1) \geq B(h-1,\,\bar h) + B(h,\,\bar h-1)
}{eq:app4}
is valid at least for $h\geq 3$ and $\bar h\geq 2$, since for every partition counted by $B(h-1,\,\bar h)$ and  $B(h,\,\bar h-1)$ there is a partition counted by $B(h,\,\bar h)$, while partitions arising simultaneously in $B(h-1,\,\bar h)$ and $B(h,\,\bar h-1)$ are counted by $B(h-1,\,\bar h-1)$.
The inequality \eqref{eq:app4} implies non-negativity of $\tilde B$, apart from the lowest order coefficients.
The multiplicity coefficients $N_{h,\,\bar h}$ in \eqref{eq:gmg29} differ from $\tilde B(h,\,\bar h)$ by some low order terms.
The explicit results in table 1 together with the combinatorial counting argument above establishes then non-negativity of all multiplicity coefficients.

\section{Conformal Chern--Simons Gravity at negative temperature}\label{app:neg}

\newcommand{\Q}{Q}

It has been suggested that CSG could make sense as a theory with negative temperature \cite{Park:2006hu} since there is an upper bound on the BTZ black hole mass.
In this appendix we calculate the 1-loop partition function for CSG at negative temperature.
Let $\tau_2$ be negative so that $|q|\geq 1$. 
Then the contributions to $\ln Z_{\rm CSG}(\tau_2<0)$, see (\ref{ZCS}), read
\begin{eqnarray}
&&-\frac 12 \ln\det (-\nabla^2-2)_2^{TT}=
-\sum_{n=1}^\infty \frac {|q|^{-2n}}{n|1-q^n|^2} (q^{2n}+\bar q^{2n})\\
&&\frac 12 \ln\det (-\nabla^2+2)_1^T=
\sum_{n=1}^\infty \frac {|q|^{-2n}}{n|1-q^n|^2} (q^{n}+\bar q^{n})\\ 
&&\ln |Z_{m=0}|=-\frac 12 
\sum_{n=1}^\infty \frac {|q|^{-n}}{n|1-q^n|^2} (q^{2n}+\bar q^{2n})\\
&&\ln Z_{\rm conf}=\sum_{n=1}^\infty \frac {|q|^{-n}}{n|1-q^n|^2}
\end{eqnarray}
Defining $\Q:=1/q$ so that $|Q|\leq 1$ and choosing the phase as in \eqref{eq:gmg13} yields then 
\eq{
Z_{\textrm{CSG}}(\tau_2<0)=1/Z_{\textrm{CSG}}(\tau_2>0)\,,
}{eq:app6}
with the CSG partition function for positive $\tau_2$ given in \eqref{eq:gmg10}, but with $q$ replaced by $\Q$.

\end{appendix}


\providecommand{\href}[2]{#2}\begingroup\raggedright\endgroup

\end{document}